\documentclass[prl,aps,showpacs,superscriptaddress,twocolumn]{revtex4}
\usepackage{graphicx}
\usepackage{amsmath}
\usepackage{amsfonts}
\usepackage{amssymb}
\usepackage{epsfig}
\usepackage{color}
\begin{document}

\newcommand\lavg{\left\langle}
\newcommand\ravg{\right\rangle}
\newcommand\ket[1]{\left|#1\right\rangle}
\newcommand\bra[1]{\left\langle#1\right|}
\newcommand\braket[2]{\left.\left\langle#1\right|#2\right\rangle}
\def\I {{\rm 1} \hspace{-1.1mm} {\rm I} \hspace{0.5mm}}
\newcommand{\rosso}[1]{\color[rgb]{0.6,0,0} #1}

\title{Protecting entanglement via the quantum Zeno effect}

\author{Sabrina Maniscalco}
\affiliation{Department of Physics, University of Turku, Turun yliopisto, FIN-20014 Turku, Finland}

\author{Francesco Francica}
\affiliation{Dip. Fisica, Universit\`a della Calabria, \& INFN -
Gruppo collegato di Cosenza, 87036 Arcavacata di Rende (CS) Italy}
\author{Rosa L. Zaffino}
\affiliation{Dip. Fisica, Universit\`a della Calabria, \& INFN -
Gruppo collegato di Cosenza, 87036 Arcavacata di Rende (CS) Italy}
\author{Nicola Lo Gullo}
\affiliation{Dip. Fisica, Universit\`a della Calabria, \& INFN -
Gruppo collegato di Cosenza, 87036 Arcavacata di Rende (CS) Italy}
\author{Francesco Plastina}
\affiliation{Dip. Fisica, Universit\`a della Calabria, \& INFN -
Gruppo collegato di Cosenza, 87036 Arcavacata di Rende (CS) Italy}
\email{sabrina.maniscalco@utu.fi}

\date{\today}

\begin{abstract}
We study the exact entanglement dynamics of two atoms in a lossy
resonator.  Besides discussing the steady-state entanglement, we
show that in the strong coupling regime the system-reservoir
correlations induce entanglement revivals and oscillations and
propose a strategy to fight against the deterioration of the
entanglement using the quantum Zeno effect.
\end{abstract}
\bigskip
 \maketitle The description of
decoherence for bipartite entangled systems has recently reached
notable theoretical \cite{yu} and experimental \cite{natu} results
due to the introduction of the concept of entanglement sudden
death. This describes the finite-time destruction of quantum
correlations due to the detrimental action of independent
environments coupled to the two subsystems. On the other hand, it
is well known \cite{palma,zanardi} that the interaction with a
{\it common} environment leads to the existence of a highly
entangled long-living decoherence-free (or sub-radiant) state. At
the same time, another entangled state exists (orthogonal to
previous one and called super-radiant) that looses its coherence
faster. In this paper, we discuss how to preserve this second
entangled state without affecting the first one by exploiting the
quantum Zeno effect in order to achieve a complete entanglement
survival.

Specifically, we address an exactly solvable non-Markovian model
describing two two-level atoms (qubits) resonantly coupled to a
lossy resonator, which we treat through the pseudo-mode approach
\cite{garraway}. As we discuss below, this describes, e.g., atoms
or ions trapped in an electromagnetic cavity \cite{Walther} or
circuit-QED setups\cite{wallra,mika,wallra2} and our results are
directly verifiable in both systems.

We obtain the exact entanglement dynamics as a function of the
environment correlation time and discuss its stationary value,
which turns out to be maximal for a factorized initial state of
the two atoms. In the past, the environment induced entanglement
generation has been discussed in the Born-Markov limit
\cite{Benatti} or for a pure dephasing case \cite{danielbraun}.
Here we extend these results to a dissipative coupling with the
environment outside the Markovian regime, both for weak and strong
couplings, corresponding to the bad and good cavity limits. In
particular, in the latter regime, the long memory of the reservoir
induces entanglement revivals and oscillations.

Furthermore, we describe a measurement induced quantum Zeno effect
\cite{zeno} for the entanglement, showing that the quite simple
procedure of monitoring the population of the cavity mode leads to
a protection of the entanglement well beyond its natural decay
time. This effect too can be tested with slight modifications of
already existing experimental set-ups, both in the realm of cavity
QED and with superconducting Josephson circuits.

We consider a two-qubits system interacting with a common
zero-temperature bosonic reservoir. The microscopic Hamiltonian of
the system plus reservoir, is given by $H=H_0 +H_{\rm int}$, with
\begin{eqnarray}
H_0 &=& \omega_1 \sigma^{(1)}_+  \sigma^{(1)}_- + \omega_2
\sigma^{(2)}_+\sigma^{(2)}_- + \sum_k \omega_k b_k^{\dag} b_k, \\
H_{\rm int} &=&  \left( \alpha_1 \sigma^{(1)}_+   +
\alpha_2\sigma^{(2)}_+ \right) \sum_k g_k b_k  + {\rm h.c.}
\end{eqnarray}
Here, $b_k$ is the annihilation operator of quanta of the
environment, while $ \sigma^{(j)}_{\pm}$ and $\omega_{j}$ are the
inversion operator and transition frequency of the $j$-th qubit,
$j=1,2$, whose interaction with the reservoir is measured by the
dimensionless constant $\alpha_j$. This depends on the value of
the cavity field at the qubit position and can be effectively
manipulated, e.g., by means of dc Stark shifts tuning the atomic
transition in and out of resonance. For the following discussion,
it will prove useful to introduce a collective coupling constant
$\alpha_T=(\alpha_1^2+\alpha_2^2)^{1/2}$ and the relative
strengths $r_j=\alpha_j/\alpha_T$ (as $r_1^2+r_2^2=1$, we take
only $r_1$ as independent). By varying $\alpha_T$, we will explore
both the weak and the strong coupling regimes.

For an initial state of the form \begin{equation}\ket{\psi(0)} =
\Bigl [ c_{01} \ket{1}_1\ket{0}_2 + c_{02} \ket{0}_1\ket{1}_2\Bigr
] \bigotimes_k \ket{0_k},\label{initialstate}\end{equation} the
time evolution of the total system is given by
\begin{eqnarray}
\vert \Psi (t) \rangle &=&  c_1 (t) \vert 1 \rangle_1 \vert 0
\rangle_2 \vert 0 \rangle_E + c_2(t)  \vert 0 \rangle_1 \vert 1
\rangle_2 \vert 0 \rangle_E+ \nonumber \\ &+& \sum_k c_k (t) \vert
0 \rangle_1 \vert 0 \rangle_2 \vert 1_k \rangle_E, \label{eq:psi}
\end{eqnarray}
$\vert 1_k \rangle_E$ being the state of the reservoir with only
one excitation in the $k$-th mode. Setting $\delta_k^{(j)} =
\omega_j-\omega_k$, the equations for the probability amplitudes
take the form
\begin{eqnarray}
\dot{c}_j(t)&=&-i \alpha_j  \sum_k g_k e^{i \delta_k^{(j)}t}
c_k(t), \quad j=1,2\label{eq:cj} \\ \dot{c}_k(t)&=&-i  g^*_k
\left[ \alpha_1 e^{-i \delta_k^{(1)} t} c_1(t) +\alpha_2 e^{-i
\delta_k^{(2)}t} c_2(t) \right]. \label{eq:ck}
\end{eqnarray}
Formally integrating Eq.~(\ref{eq:ck}) and inserting its solution
into Eqs.~(\ref{eq:cj}), one obtains two integro-differential
equations for the amplitudes $c_{1,2}(t)$. In the continuum limit
for the environment, and with the introduction of the correlation
function $f(\tau)$, and  of its Fourier transform $J(\omega)$
(which is the reservoir spectral density), these two equations
become
\begin{eqnarray}
&&\dot{c}_1=- \int_0^t \! dt_1 f(t-t_1) \left[ \alpha_1^2
c_1(t_1) +
\alpha_1 \alpha_2 c_2(t_1) \right], \label{eq:c1int} \\
&&\dot{c}_2=- \int_0^t \! dt_1  f(t-t_1)  \left[ \alpha_2^2
c_2(t_1) + \alpha_2 \alpha_1 c_1(t_1) \right]. \label{eq:c2int}
\end{eqnarray}
Before discussing the general time evolution, we notice that a
constant solution can be found, independently of the form of the
spectral density. Namely, a sub-radiant, decoherence-free state
exists, that does not decay in time
\begin{eqnarray}
\label{eq:psim} \vert \psi_- \rangle = r_2 \vert 1\rangle_1 \vert
0 \rangle_2  - r_1 \vert 0\rangle_1 \vert 1 \rangle_2.
\end{eqnarray}
In the following, we consider the case in  which the two atoms
have the same Bohr frequency, i.e., $\omega_1 = \omega_2 =
\omega_0$, and interact resonantly with a reservoir with
Lorentzian spectral density
\begin{eqnarray}
\label{eq:J} J(\omega) = \frac{W^2}{\pi} \frac{\lambda}{\left(
\omega - \omega_0 \right)^2 + \lambda^2}.
\end{eqnarray}
This is, e.g., the case of two atoms interacting with a cavity
field in presence of cavity losses. Due to the non perfect
reflectivity of the cavity mirrors, the spectrum of the cavity
field displays a Lorentzian broadening. In this case the
correlation function decays exponentially $f(\tau)= W^2
e^{-\lambda \tau}$,  the quantity $1/ \lambda$ being the reservoir
correlation time. The ideal cavity limit is obtained for $\lambda
\rightarrow 0$; in this case one has
\begin{equation}
\label{ eq:Jideal} \lim_{\lambda \rightarrow 0} J(\omega) = W^2
\delta(\omega - \omega_0),
\end{equation}
corresponding to a constant correlation function $f(\tau)=W^2$.
The system, then, reduces to a two-atom Jaynes-Cummings
model \cite{tavis} with a vacuum Rabi frequency  ${\cal R}=
\alpha_T W$. On the other hand, for small correlation times (with
$\lambda$ much larger than any other frequency scale), we obtain
the Markovian regime characterized by a decay rate $\gamma = 2
{\cal R}^2/\lambda$. For generic parameter values, the model
interpolates between these two limits.

As $\ket{\psi_-}$ does not evolve in time, the only relevant time
evolution is that of its orthogonal, super-radiant, state
\begin{eqnarray}
\label{eq:psip} \vert \psi_+ \rangle = r_1 \vert 1\rangle_1 \vert
0 \rangle_2  + r_2 \vert 0\rangle_1 \vert 1 \rangle_2.
\end{eqnarray}
One obtains $\vert \psi_+ (t) \rangle = {\cal E}(t) \vert \psi_+
(0) \rangle$, with
\begin{eqnarray}
{\cal E}(t)  = e^{- \lambda t /2} \left[ \cosh \left ( \Omega t
/2\right) + \frac{\lambda}{\Omega} \sinh \left( \Omega t /2
\right) \right], \label{eq:E}
\end{eqnarray}
where $\Omega= \sqrt{\lambda^2 -  4 {\cal R}^2}$.

In the $\{\vert 1\rangle, \vert 0 \rangle \}$ basis, the reduced
density operator for the two qubits is given by
\begin{equation}
\label{eq:rhos} \rho(t) = \begin{pmatrix}
   0& 0 & 0 & 0 & \\ 0& |c_1(t)|^2   &  c_1(t)c_2^*(t)  & 0\\
 0& c_1^*(t)c_2(t)      & |c_2(t)|^2  & 0\\
   0&  0  & 0 & 1-|c_1|^2-|c_2|^2
\end{pmatrix} ,
\end{equation}
where, letting $\beta_{\pm}= \braket{\psi_{\pm}}{\psi(0)}$, one
has
\begin{eqnarray}
c_1(t) &=& r_2 \, \beta_-
+ r_1 \, {\cal E}(t) \,  \beta_+, \label{eq:c1t} \\
c_2(t) &=&- r_1 \, \beta_- + r_2 \, {\cal E}(t)\, \beta_+.
\label{eq:c2t}
\end{eqnarray}
The solution is exact as we have not performed neither the Born
nor the Markov approximation. We now use this result to obtain
the dynamics of the qubit entanglement as measured by the
concurrence $C(t)$ \cite{wootte}, ranging from 0 (for separable
states) to 1 (for maximally entangled ones). For the density
matrix given by Eq.~(\ref{eq:rhos}), the concurrence is
\begin{equation}
\label{eq:cdef} C(t) = 2 \left| c_1(t) c_2^*(t) \right| .
\end{equation}

\begin{figure}
\includegraphics[width=.45\textwidth]{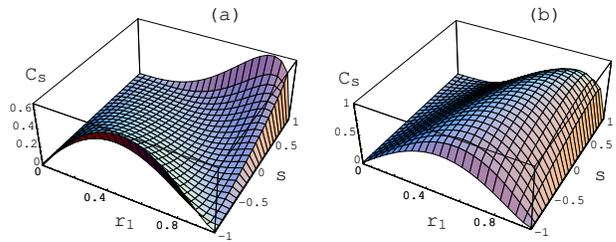}
\caption{(Color online) Stationary concurrence as a function of
the relative coupling constant $r_1$ and of the initial
separability $s$ of the state,  for (a) $\phi=0$, and (b)
$\phi=\pi$. In the first case, the maximum of $C_s$ is achieved
for asymmetrical couplings: at $r_1 \simeq 0.87$ for $s=1$, and at
$r_1 \simeq 0.5$ for $s=-1$. In the second case the maximum is
achieved in correspondence of $\ket{\psi(0)}=\ket{\psi_-}$.}
\label{fig:csta}
\end{figure}

\begin{figure}
%\begin{center}
\includegraphics[width=.50\textwidth]{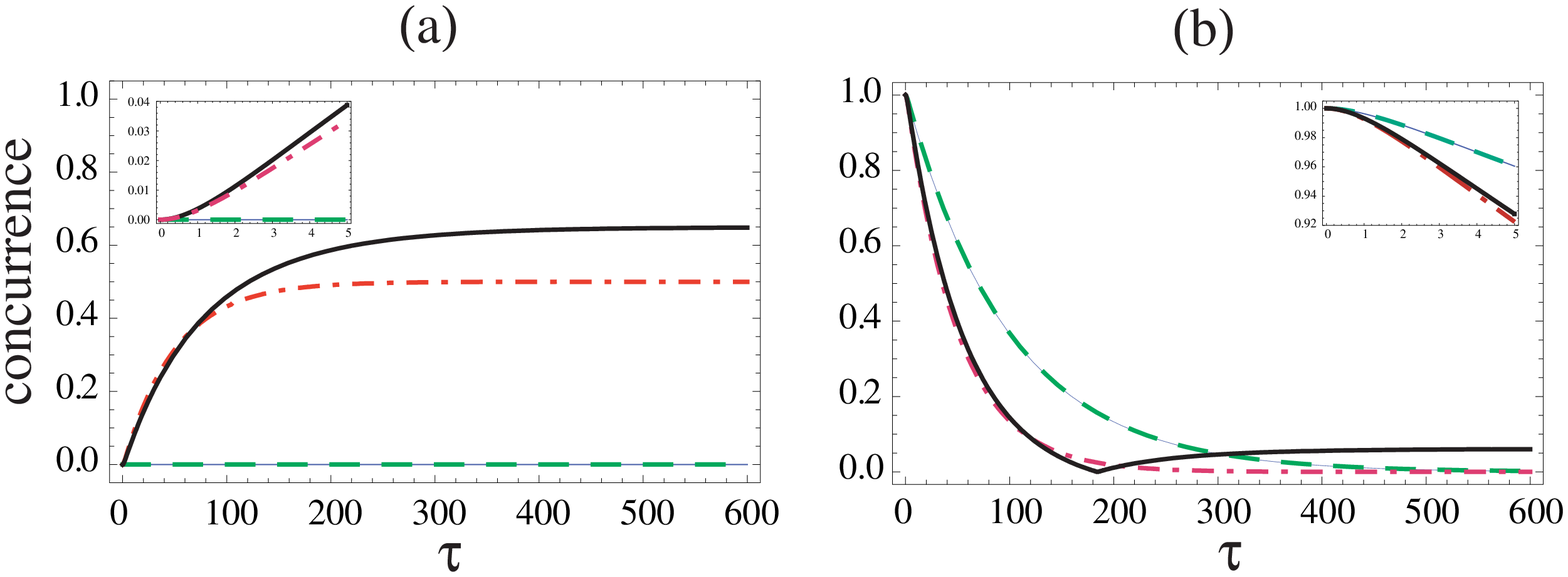}
\includegraphics[width=.50\textwidth]{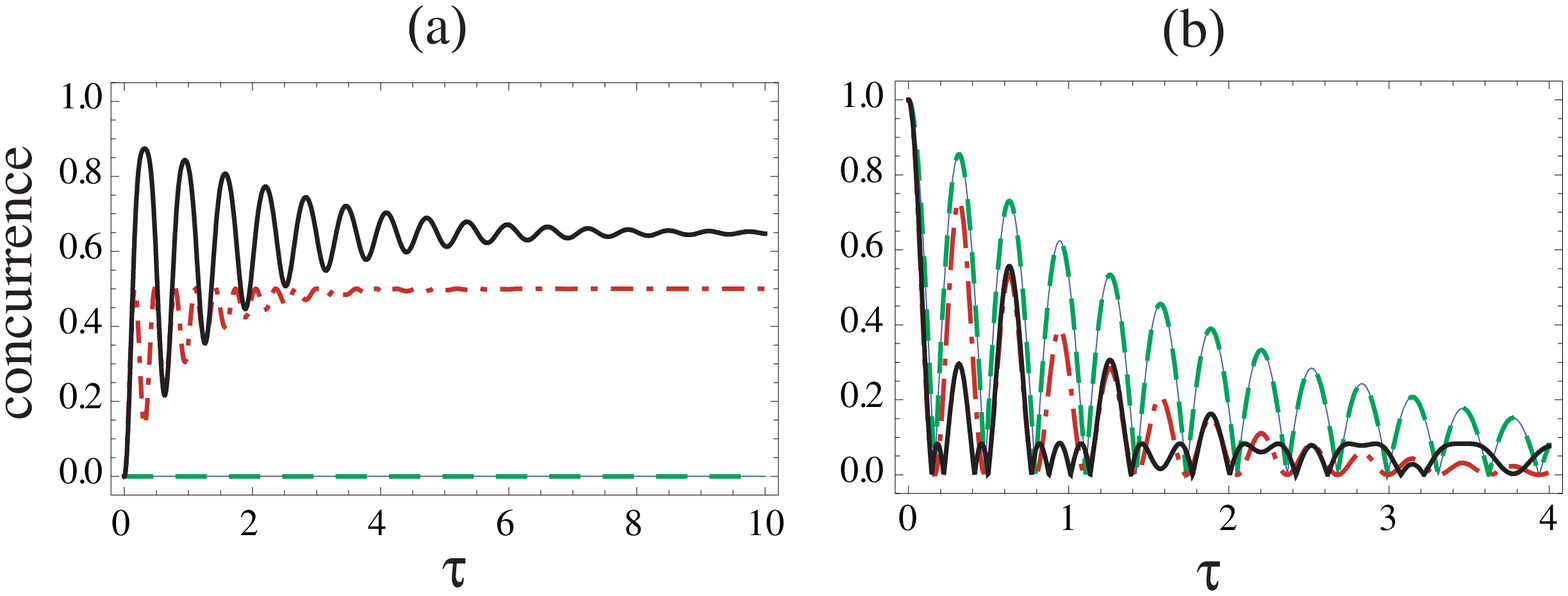}
\caption{(Color online) Time evolution of the  concurrence in the
bad cavity limit ($R=0.1$, top plots) and good cavity limit
($R=10$, bottom plots), with (a) $s=1$, and (b) $s=0$, both with
$\phi=0$, for the cases of i) maximal stationary value, $r_1=0.87$
(solid line), ii) symmetrical coupling $r_1=1/\sqrt{2} $
(dot-dashed line), and iii) only one coupled atom $r_1=0,1$
(dotted line).  The insets show the initial quadratic behavior of
the concurrence for $R=0.1$} \label{fig:weakstrongtot}
%\end{center}
\end{figure}
We begin by noticing that there exist a non-zero stationary value
of $C$ due to the entanglement of the decoherence-free state: $C_s
= C(t \rightarrow \infty) \equiv C(\ket{\psi_-}) \, \left
|\braket{\psi_-}{\psi(0)} \right |^2 = 2 |r_1 r_2| \, \left|
\beta_- \right|^2$. To better discuss the time evolution of the
concurrence as a function of the initial amount of entanglement
stored in the system, we consider initial states of the form
(\ref{initialstate}) with
$$c_{01}= \sqrt{\frac{1-s}{2}}, \quad c_{02}= \sqrt{\frac{1+s}{2}} \,
e^{i \phi}, \mbox{with} -1 \le s \le 1. $$ Here, the separability
parameter $s$ is related to the initial concurrence as
$s^2=1-C(0)^2$. Fig.~\ref{fig:csta}-(a) displays the stationary
concurrence versus $r_1$ and $s$. Due to the interaction with the
cavity field, initial separable states $(s=\pm 1)$ become
entangled. In fact, for $\phi=0$, the maximum stationary
entanglement $C_s^{\rm max} \simeq 0.65$ is obtained for initially
factorized states, i.e. $s=\pm 1$. While the details depend on the
phase $\phi$, the qualitative picture is generic and essentially
independent of $\phi$, apart from the isolated case of an initial
state coinciding with $\ket{\psi_-}$. In such a situation all of
the entanglement initially encoded in the qubits remains there for
long times. For positive $r_j$, this occurs for $\phi=\pi$, see
Fig. \ref{fig:csta}-(b).

We now look at the entanglement dynamics  in the good and bad
cavity limits, i.e. for $R\gg 1$ and $R \ll 1$, respectively, with
$R= \mathcal{R}/\lambda$.  In Fig.~\ref{fig:weakstrongtot} we show
the concurrence as a function of $\tau = \lambda t$ in the bad
(upper row) and good (lower row) cavity limits. We compare the
dynamics of an initially  factorized state ($s=1$) with the one of
an initially maximal entangled state ($s=0$) for four different
values of the coupling parameter $r_1$, namely
$r_1=0,1/\sqrt{2},0.87,1$. The plots for $r_1=0$ and $r_1=1$
overlap as they both describe a case in which one of the two atoms
is effectively decoupled. The value $r_1=0.87$ corresponds to the
case of optimal stationary entanglement for the initial state
$s=1$ with $\phi=0$. Finally  $r_1=1/\sqrt{2}$ corresponds
to the case of equal coupling of the two atoms with the reservoir.
Other values of $r_1$ show qualitatively similar behavior.

For weak couplings and/or bad cavity, $R=0.1$, and for an
initially separable state ($s=1$), the concurrence increases
monotonically up to its stationary value; whereas, for initially
entangled states, the concurrence first goes to zero before
increasing towards $C_s$. The strong coupling/good cavity case
$R=10$ is more rich and presents entanglement oscillations and
revival phenomena for all the initial atomic states. One can
prove analytically that for maximally entangled initial states
($s=0$) the revivals have maximum amplitude when only one of the
two atoms is effectively coupled to the cavity field, i.e. for
$r_1=0,1$. In this case, indeed, the system performs damped
oscillations between the states $\vert \psi_+ \rangle$ and $\vert
\psi_- \rangle$ which are equally populated at the beginning. On
the other hand, for an initially factorized state, the interaction
with the cavity field in the strong coupling regime generates a
high degree of entanglement. Indeed, for $R=10$,  at $\tau =
\bar{\tau} \simeq 0.31$,  $C$ attains the value
$C(\bar{\tau})\simeq 0.96$, at $r_1 \simeq 0.92$ (for $s=1$) or
$r_1 \simeq 0.4$ (for $s=-1$) .

These entanglement revivals are a truly new
result due to the memory depth of the reservoir. Very small revivals can occur in the Markovian
regime \cite{ficek}, and in the non-Markovian regime with
independent reservoirs \cite{bellomo}. In our case, however, the amount of revived entanglement is huge, being
comparable to the previous maximum. This  feature only appears in
the strong coupling regime and with a non-zero environmental
correlation time. The surprising aspect, here, is that an
irreversible process  such as the spontaneous emission is so
strongly counteracted by the memory effect of the environment,
which not only creates entanglement, but also lets it oscillate
quite a few times before a stationary value is reached.

If we express the initial state of the qubits as a superposition
of $\vert \psi_{\pm} \rangle$, that is $\vert \psi (0) \rangle =
\beta_- \vert \psi_- \rangle + \beta_+ \vert \psi_+ \rangle$, we
see that, while part of the initial state will be trapped in the
sub-radiant state $\vert \psi_- \rangle$, another part will decay
following Eq.~(\ref{eq:E}). Thus, as discussed above, the amount
of entanglement that survives depends on the specific state (and
on the value of the $r_j$). In the following we present a method
which exploits the quantum Zeno effect to preserve the initial
entanglement {\it independently} of the state in which it is
stored.

We consider the action of a series of nonselective measurements on
the collective atomic system, performed at time intervals $T$,
which have the two following properties: i) one of the possible
measurement outcomes is the projection onto the collective ground
state $\vert \psi_g \rangle = \vert 0 \rangle_1 \vert 0
\rangle_2$, and ii) the measurement cannot distinguish between the
states $\ket{1}_1 \ket{0}_2$ and $\ket{0}_1\ket{1}_2$. Any
procedure fulfilling these two conditions will do the job of
preserving the entanglement. In particular, one could measure the
collective atomic energy [condition ii) holds in this case since
the transition frequencies are equal] or, more simply, one could
monitor the state of the cavity: if a photon is found, than the
qubits have necessarily decayed into $\ket{\psi_g}$, while if no
photon is found, than the excitation still resides on the atoms.
This can be done both in cavity QED set-ups (by sending a probe
atom through the cavity that can absorb the photon) and with
superconducting circuits (by sending a short measuring voltage
pulse to the resonator, similarly to Ref. \cite{wallra2}).

The measurements described above disentangle the qubits from the
reservoir at each time $T$. Choosing $T$ such that  $\langle
\Psi_g \vert \rho (T) \vert \Psi_g \rangle \ll 1$, it is
straightforward to prove that the state $\ket{\psi_-}$ is
unaffected and that, at the same time, the decay of $\ket{\psi_+}$
is slowed down. Its survival probability $P_+^{(N)}(t) = \langle
\psi _+ \vert \rho (t) \vert \psi_+ \rangle$ in presence of $N$
measurements is given by $P_+^{(N)}(t) =\vert \beta_+(0) \vert^2
\exp \left[ - \gamma_z(T)t \right]$, where $t=N T$ and with an
effective decay rate
\begin{equation}\gamma_z(T) = - \frac{\log \left[ \mathcal{E}(T)^2
\right]}{T}.\end{equation} Notice that, in the limit $T
\rightarrow 0$ and $N \rightarrow \infty$, with a finite $t=NT$ ,
$\gamma_z(T) \rightarrow 0$ and the decay is completely
suppressed.

\begin{figure}
\includegraphics[width=.50\textwidth]{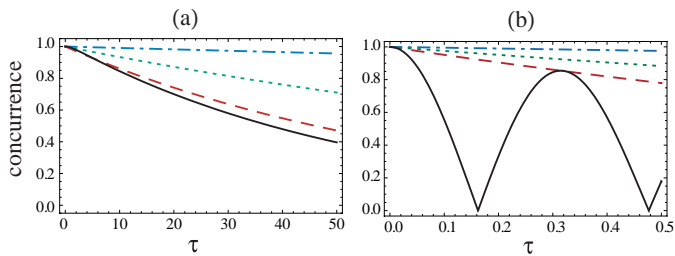}
\caption{Time evolution of the concurrence, for $s=0$ and
$r=1/\sqrt 2$, in absence of measurements  (solid line) and in
presence of measurements performed at intervals (a) $\lambda
T=0.1,1,5$ (dashed, dotted  and dot-dashed lines, respectively)
with $R=0.1$ (weak coupling) and (b) $\lambda T=0.001,0.005,0.01$
(dashed, dotted, and dot-dashed lines, respectively) with $R=10$
(strong coupling).} \label{fig:zeno}
\end{figure}
Besides affecting the probability $P_+(t)$, the projective
measurements also modify  the time evolution of the entanglement,
whose effective dynamics now depends on $T$. Explicitly, the
concurrence at time $t=N T$, after performing $N$ measurements, is
given by
\begin{eqnarray}
C^{(N)}(t) = 2 & \Bigl |& \left (\beta_+ r_1 \, e^{- \gamma_z t/2}
+ \beta_- r_2 \right )  \times \nonumber \\ && \left(\beta_+ r_2
\, e^{- \gamma_z t/2} - \beta_- r_1 \right ) \Bigr |.
\end{eqnarray}
In Fig.~\ref{fig:zeno} we compare the dynamics of $C(\tau)$ in
absence  and in presence of measurements performed at various
intervals $T$ for an initially maximal entangled state. Both in
the weak and in the strong coupling regimes (left and right plots,
respectively) the presence of measurements  quenches the decay of
the concurrence. Thus, we have achieved a quantum Zeno protection
of entanglement from the effect of decoherence. Again, decreasing
the interval between the measurements, $C^{(N)}(t)$ remains closer
and closer to its initial value.

To sum up, we discussed an exactly solvable model describing two
qubits interacting with a non ideal resonator. We analyzed in
detail the stationary entanglement and obtained the entanglement
dynamics both in the weak and strong coupling limits, showing that
entanglement revivals can appear due to the finite memory of such
a complex environment. We also investigated the quantum Zeno
effect for this system, showing that the entanglement can be
preserved independently of the state in which it is encoded, with
the help of repeated projective measurements. As anticipated above, our results apply both to cavity
QED experiments with trapped atoms and to the case of
superconducting circuits, with on-chip qubits and resonator. In
the first case, it has already been demonstrated that both atoms
and ions can be confined inside high finesse optical cavities and
their quantum states can be fully controlled \cite{Walther}. In
the second case, quantum communication between two Josephson
qubits has been achieved using a transmission line as a cavity
\cite{mika,wallra2}.  

S.M. thanks Francesco Plastina and his group for the kind hospitality at the Universit\`a della Calabria and acknowledges financial support from the Academy of Finland (projects 108699, 115682), the Magnus Ehrnrooth Foundation and the V\"{a}is\"{a}l\"{a} Foundation.

\end{document}